# Two-Phase Equilibrium in Small Alloy Particles

J. Weissmüller,[#] P. Bunzel, G. Wilde

*Institut für Nanotechnologie, Forschungszentrum Karlsruhe,
and Technische Physik, Universität des Saarlandes*

**Keywords:** nanocomposite; phase transformations; thermodynamics

**Abstract**

The coexistence of two phases within a particle requires an internal interface with a significant capillary energy. We show that this entails changes in the nature of alloy phase equilibria at small size. Most notably, the eutectic points in alloy phase diagrams degenerate into intervals of composition where the alloy melts discontinuously.

**1. Introduction**

When the size of a particle is reduced then the excess free energy due to the surface diminishes more slowly than the free energies of the bulk phases, and capillary effects will therefore increasingly affect the thermodynamic equilibrium. The consequences of this phenomenon have been studied for more than a century. Yet, size-dependent phase equilibria have retained their fascination, as exemplified by many contemporary studies of melting of *elemental* nanoparticles.[1,2,3,4,5] The interest in phase equilibria of nanoscale *alloys* is more recent. Grain boundary segregation and its effect on the equilibrium, in particular through changes in the interface stress, is among the subjects of interest, for instance in nanocrystalline metal hydrides.[6] Recently, studies have turned to phase transitions in matrix-embedded alloy nanoparticles, prepared by vapor deposition,[7] ion implantation,[4] or by melt spinning.[8,9] As the temperature is cycled, the number and the nature of the coexisting phases in each particle change, and this requires interfaces which separate the phases to be created or removed within the particles. The new issue is, how does the energetics of this process affect the phase equilibrium? Here, we show that alloy phase diagrams exhibit qualitative changes at small particle size. These are not mere shifts of temperatures or of compositions at equilibrium; instead, rules which universally apply to alloy phase diagrams in the limit of macroscopic systems size fail at the nanometre scale.

**2. Free Energy Functions**

We analyze two-phase equilibria in a particle, or in sets of identical particles, with fixed amounts $N_1$ and $N_2$ of solvent and solute, respectively, in each particle. In a typical experiment, the particles are embedded in a solid or fluid matrix which serves to suppress the exchange of matter between them, thereby preventing coarsening and changes of composition. Thus, as the temperature is cycled, each particle may undergo reversible phase changes.

The situation is similar to that considered in the classical theory of nucleation: small volumes of one phase (here: the particles themselves, or even smaller objects, namely precipitates of a second phase within each particle) are embedded in another phase, and the interfacial terms are relevant. However, nucleation theory considers unstable transient states created by thermal fluctuations, nuclei of a new phase which exchange matter with a macroscopic reservoir, the parent phase. By contrast, the systems considered here are closed, and at any given set of values for $T$, $N_1$, $N_2$ they have a well-defined equilibrium state. Transitions between such states may require that an activation barrier is overcome by nucleation (*within* the particle), but the focus of our interest is on the final equilibrium states, rather than on nucleation.

Let us denote the two phases by superscripts α and β, and let the functions $G^\alpha(T,N,x)$ and $G^\beta(T,N,x)$ represent the free energies, per particle, of single-phase states containing the fixed amount $N = N_1 + N_2$ of matter per particle (Fig. 1a). $T$ denotes the temperature and $x = N_2/N$ the solute fraction. In the limiting case of very large (macroscopic) particles $G^\alpha$ and $G^\beta$ are homogeneous first order functions of $N$, $G(T,N,x) = N\,g(T,x)$ where $g$ denotes the molar free energy. At finite size the free energy of the external surface must be considered. In the simplest case - fluid particle embedded in a fluid matrix - the total superficial free energy is $\gamma A$ with $\gamma$ a specific surface free energy and $A$ the surface area. Crystalline solids exhibit various types of interfaces distinguished by their crystallography and composition. The free energy per particle can then be expressed by

$$G(T,N,x) = N\,g(T,x) + \Sigma_i\,\gamma_i(T)\,A_i(T,N,x) \qquad (1)$$

where the subscript labels the possible interfaces. At equilibrium, the $A_i$ are not independent state variables, but internal thermodynamic parameters which are functions of $T,N,x$, determined by the Wulf construction. The appropriate constitutive equation is therefore $G=G(T,N,x)$, rather than $G=G(T,N,x,A_i)$. Generally the functional dependence of the $A_i$ on $N$ is not linear; this leads to the size-dependence of the chemical potentials of single-phase particles embodied in Gibbs-Thompson-

---

[#]: Corresponding author. E-Mail Joerg.Weissmueller@int.fzk.de

Freundlich type equations.

Consider now the case where each particle contains two coexisting phases. In the macroscopic case the total free energy is a linear function of the molar phase fraction. According to the lever rule, the fraction of phase α, $p^\alpha = N^\alpha/N$, obeys

$$p^\alpha(x, x^\alpha, x^\beta) = (x^\beta - x) / (x^\beta - x^\alpha). \quad (2)$$

We denote by $\tilde{G}$ the free energy when two phases of arbitrary compositions, which may not be the compositions at equilibrium, coexist. In the macroscopic system

$$\tilde{G}(T, N, x, x^\alpha, x^\beta) = p^\alpha G^\alpha(T, N, x^\alpha) + (1-p^\alpha) G^\beta(T, N, x^\beta). \quad (3)$$

This linear dependence of $\tilde{G}$ on $p$ provides the basis for the common tangent construction by which the compositions of the phases coexisting at equilibrium may be found.

The nucleation and growth of a new phase in a particle will entail changes of the area of the interfaces, and the creation of new interfaces which separate the new phase from the matrix and/or the parent phase. In addition, the equilibrium shape of the particle may change, so that the crystallography as well as the area of the interfaces vary. In general, the dependency of $\tilde{G}$ on the phase fraction will then not be linear, contrary to Eq. (2), and we may represent this deviation from linearity by a term $\Delta G_C$, defined so that

$$\tilde{G}(T, N, x, x^\alpha, x^\beta) = p^\alpha G^\alpha(T, N, x^\alpha) + (1-p^\alpha) G^\beta(T, N, x^\beta) + \Delta G_C. \quad (4)$$

In the schematic graph of Fig. 1a) the free-energy curve for the two-phase state joins the straight line which represents the linear behavior in the points A and B representing the compositions of the coexisting phases since, by definition, $\Delta G_C = 0$ for single-phase particles. In between, a curved graph must result whenever $\Delta G_C$ has nonvanishing values.

For the purpose of illustration schematic cross-sections of the particles are included in Fig. 1a). They refer to the example of spherical particles in which the internal interface meets the particle surface at right angles. The free energy graph is convex since the area and, therefore, the energy of the internal interface pass through a maximum when both phases have equal volumes. Other geometries are possible, depending on the relative values of the $\gamma_i$ (compare Ref. 8); they may give rise to different functional forms of $\Delta G_C$, but quite generally $\Delta G_C$ is a nonlinear function of the phase fraction. The important consequence of this loss of linearity is that the tangent rule ceases to apply. Instead, the free energy at equilibrium, $G^{\alpha\beta}(T, N, x)$, is the lower envelope of the set of functions $\tilde{G}^{\alpha\beta}(T, N, x, x^\alpha, x^\beta)$. Well-known consequences of the common tangent construction, which must generally be obeyed by all macroscopic alloy systems, are that $x^\alpha$ and $x^\beta$ are invariant when $x$ is varied, and that the composition of the majority phase is continuous across phase boundary lines. As the common tangent construction breaks down in small systems, it is of interest to verify if its consequences continue to hold.

It is noted that the situation has analogs to the equilibrium between coherent phases, where the coherency strain energy is described by convex free energy functions similar to Fig. 1a).[10,11,12] It is also worth mentioning that single-phase states with convex free energy functions would be unstable with respect to the formation of a two-phase state by spinodal decomposition, but that there is no such instability here since two-phase states are considered from the outset.

### 3. Size-dependent Alloy Phase Diagram

Numerical computation was necessary for constructing alloy phase diagrams for finite size systems. We study an exemplary case using simple equations of state. In a reduced representation the three materials constants which need to be specified take on similar values for most metals. By using these values we achieve results which can be compared to experiments in a semi-quantitative way.

We consider an alloy with no solid solubility and with an ideal solution as the melt. The phase diagram of the macroscopic alloy is of the simple eutectic type, Fig. 2a. For simplicity we assume identical temperatures of fusion $T_f$ for the pure components; the free energy functions are then invariant with respect to replacement of solute by solvent, and the phase diagram is symmetric about $x = \frac{1}{2}$. We use the identical molar volume, $v$, for all components in all phases, so that the volume of the particle will not change, and Clausius-Clapeyron type, pressure-induced shifts of phase equilibria can be ignored.

In order to achieve a simple geometry, we fix the particle shape to be spherical, and we take all interfacial free energies to have a constant and identical value, $\gamma$. This has three important consequences: *i*) changes in the excess free energy arise exclusively from the variation of the internal interface area, $\Delta G_C = \gamma A$; *ii*), there is no solute segregation, and *iii*), the dihedral angle at the junction line of the internal interface with the outer surface of the particle is 90°. The internal interface will then be a spherical cap, similar to Fig. 1a.

At fixed $N$ there are four possible states and four corresponding free energy functions: single-phase liquid, $G^L(T, x)$, two-phase pure solid '1' plus pure solid '2', $G^{SS}(T, x)$, and the two solid-liquid states. Since the phase diagram is symmetric, it is sufficient to consider only one of the solid-liquid equilibria, without lack of generality solid solvent plus liquid, $\tilde{G}^{SL}(T, x, x^L)$.

It is conceivable that more than two phases coexist at equilibrium. However, all three-phase configurations which we examined had larger interface area and, hence, larger free energy, than the competing two-phase states. Therefore, three-phase coexistence was not considered further in the computation. What is more, according to the phase rule[13] any given phase can only coexist with two others at *points* represented by discrete combinations of $T$ and $x$ in its phase field in binary alloy phase diagrams such as Fig. 2. Thus, even if there is a stable three-phase state (as in a macroscopic alloy), the single-phase and two-phase fields would still represent the equilibrium phase diagram correctly everywhere except in a single point in ($T$, $x$)-space, and the lines delimiting the phase fields in the alloy phase diagram would extrapolate to that point.

Adding a constant to the chemical potentials of any one of the components in all phases does not change the equilibrium states. Therefore, we can define the free energies of the elemental solid particles at any given temperature $T$ as zero, $G^{S_1}(T) = G^{S_1}(T) \equiv 0$. The free energies of the two-phase states are then $G^{SS} = \gamma A$ and $G^{SL} = (1-p)Ng^L + \gamma A$, where $p$ denotes the phase fraction of solid solvent.

Figure 1c) shows an example of the free energy functions analyzed in the computation (compare Appendix). The red and blue lines denote, respectively, the single-phase liquid and the solid-solid coexistence. The green lines denote the set of free energy curves for solid-liquid coexistence, $\tilde{G}^{SL}(T,x,x^L)$, with the solute fraction $x^L$ in the liquid as a parameter. To find the phase diagram, we have first computed the envelope function $G^{SL}(T,x)$ (black line) by identifying the values $x^L_0(T,x)$ of the solute fraction in the liquid which minimize $\tilde{G}^{SL}(T,x,x^L)$ at any given combination of $x$ and $T$. By construction, $G^{SL}(T,x)$ represents the equilibrium state of the particle if it is constrained to be in a two-phase solid-liquid state, and $x^L_0(T,x)$ represents the composition of the liquid at equilibrium. We have then identified the state (liquid, solid-liquid, or solid-solid) of minimum free energy by comparing the values of the functions $G^L$, $G^{SL}$, and $G^{SS}$. It is emphasized that all possible states of the alloy are investigated, and that therefore the minima identified in the computation represent *absolute* minima of the free energy at the given values of $T$, $x$ and $N$.

Figures 2a-c) show the phase diagrams of the macroscopic system and of nanoparticles with diameters $D = 50$ nm and $D = 5$ nm, respectively. It is seen that, as the particle size is reduced, the phase diagram undergoes several qualitative changes, each of which breaks one of the rules which apply universally to the construction of the phase diagram for macroscopic systems. First, it is observed that the invariance of the solidus temperature is lost in favor of a significant composition-dependence. Second, as illustrated by the lines representing states of identical composition $x^L$ of the liquid phase at equilibrium (colored lines), the compositions of the constituent phases in two-phase equilibria are no longer invariant at constant $T$. Thirdly, the equi-composition lines lose their continuity at the intersection with the liquidus line. This implies that there is a discrete jump in liquid fraction across the liquidus of the small alloy particles, consistent with the result of numerical modeling matched to Sn-Bi nanoparticles,[14] where the ends of the tie lines were found to detach from the phase boundary lines.

However, the most fundamental consequence of the finite system size is a topological change in the phase diagram, the degeneration of the eutectic point of the macroscopic system into a line representing an interval of compositions $\Delta x_d$ (defined in Fig. 2c)) for which the particle undergoes a discontinuous transition between the two-phase solid-solid state and the single-phase liquid state. In the macroscopic system, three phases can coexist at equilibrium at the eutectic point; by contrast, discontinuous melting in our model is a transition between a 2-phase equilibrium (solid-solid) and a single phase state, without three phase equilibrium. It is because of this loss of three-phase equilibrium in the finite-size system that the transition from eutectic *point* to discontinuous melting *line* can be reconciled with the phase rule.

We found numerically that, asymptotically at large $D$, the discontinuous melting interval $\Delta x_d$ in the model alloy varies as $D^{-3/4}$. The next section discusses discontinuous melting and the exponent of the size-dependence.

## 4. Discontinuous Melting Interval

For a traceable derivation we investigate large (but finite-sized) particles, restricting attention to alloys with $x$ very close to the solute fraction, $x_e$, of the macroscopic eutectic alloy. A reasonably general expression for $\Delta G_C$ can then be found, which allows an approximate closed form solution. Conclusions on the nanoscale systems may be obtained by extrapolation to small system size.

With $x$ close to $x_e$, the solid phase fraction in solid-liquid equilibria can only be small. Irrespective of whether the solid forms in the interior of the liquid particle, at its surface, or in the matrix adjacent to it, it can be treated as a precipitate much smaller than the particle. A fourth geometry, wetting of the surface by the solid, will be disregarded here since the absence of a nucleation barrier for melting of metals indicates the opposite case, premelting of the surface and, hence, wetting of the particle surface by the liquid.[1,15]

For isotropic $\gamma$ the interfaces are curved and, in the limit of small solid fraction, the local radii of curvature $r^{\alpha\beta}$ are much smaller for the precipitate surfaces compared to the particle surface. If a small amount of matter is added or removed at a surface, then the local change in area $\delta A$ is related to the volume change $\delta V$ by[16] $\delta A = 2\,\delta V/r$. Thus, when matter is transferred between the phases at small solid fraction, the change in particle surface area is negligible compared to that of the precipitate surface area. This statement holds even when the molar volume changes during the transformation, and it remains valid in the case of anisotropic $\gamma$, since the total surface area of faceted particles is comparable to that of spherical ones of identical volume. As the precipitate grows, its equilibrium shape is determined by the specific interfacial free energies, which are independent of its size. The growth will then be by affine scaling, which changes all surface areas and, hence, the total surface free energy, proportional to the precipitate volume $V^S$ to the power of 2/3. The same holds true for that region of the original liquid-matrix interface which is consumed by a growing precipitate. At small $p$, the variation of the molar interfacial energy due to precipitate growth will therefore obey

$$\Delta G^C = \Gamma (V^S)^{2/3} \qquad (5)$$

where $\Gamma$ is a constant which depends on the precipitate geometry and on the energetics of the various interfaces, but which has the same units and magnitude as a specific interfacial free energy. For the examples of (*i*) a spherical and (*ii*) a cubic precipitate in the interior of the particle, and of (*iii*) the hemispherical precipitate in the numerical model, the values of $\Gamma$ are $(36\pi)^{1/3}\gamma^{SL} \approx 4.8\,\gamma^{SL}$, $6\gamma^{SL}$, and $(18\pi)^{1/3}\gamma^{SL} \approx 3.8\,\gamma^{SL}$ respecti-vely.

Figure 1b) is a schematic representation of the free energy functions at $T$ slightly above the eutectic temperature of the particle. As a consequence of Eq. (5) the free energy curve $G^{SL}$ of the two-phase solid-liquid state meets that of the single phase liquid, $G^L$, with a diverging slope, which gives rise to an interval of $x$ where solid and liquid cannot coexist since $G^{SL} > G^L$. In the figure, the single phase liquid state is stable in a small composition interval around the eutectic point, and the two phase solid-solid state is stable otherwise. Due to the higher entropy of the liquid, its free energy curve will shift to lower free energy, relative to that of the solid, as $T$ is increased, and the composition interval for stability of the melt will widen accordingly. Alloys with a composition within this interval melt discontinuously. The trend for widening the interval of stability of the melt at the expense of the two-phase solid-solid state with increasing $T$ will continue until, at a sufficiently high temperature $T_d$, the point of intersection between $G^L$ and $G^{SL}$ (at $x = x_d$) meets the curve $G^{SS}$. Only beyond $T_d$ can there be an interval of composition in which two-phase solid-liquid states are stable.

The width of the discontinuous melting interval is given by the value of $x_d$ at the intersection of the three free energy curves $G^L$, $G^{SS}$, and $G^{SL}$ (see Fig. 1b) at the temperature $T_d$. An estimate for $x_d$ is derived in the Appendix. In symmetric alloys $\Delta x_d = 2(x^L_e - x_d)$; the result for $x_d$ then implies

$$\Delta x_d = 4 \left( \frac{3}{\pi (x^L_e - x^S_e)^2} \right)^{1/4} \left( \frac{\Gamma v^S \chi^L_e}{D} \right)^{3/4}. \qquad (6)$$

It is seen, that the theory reproduces the particle-size exponent found in the numerical computation.

## 5. Concluding Remarks

Several of the rules that apply generally to the construction of phase diagrams for macroscopic alloy systems are violated at small particle size. Most notably, a discontinuous melting interval $\Delta x_d$ near eutectic points appears at small system size. In the example, which uses materials parameters characteristic for 'typical' metals, $\Delta x_d$ reaches values in excess of a few atomic percent at particle sizes of roughly 100 nm and below. This work was supported by DFG (SPP 1120 and Center for Functional Nanostructures).

## Appendix:
### Details of the phase diagram computation

*Molar free energy $g^L$ of the melt:* For the ideal solution, $g^L = \Delta h_f - T(\Delta s_f + \Delta s_{mix})$, where $\Delta s_{mix}$ is the entropy of mixing of a random solution. $\Delta h_f$ and $\Delta s_f$ denote the enthalpy and entropy of fusion, respectively, of the pure components, assumed identical for both components. When all molar energies are measured in units of $\Delta h_f$, then $g^L$ can be expressed in terms of the dimensionless parameters $T/T_f$ and $\Delta s_f/R$ ($R$ - gas constant) by $g^L = 1 + (T/T_f)(-1 + [R/\Delta s_f(x \ln x + [1-x]\ln[1-x])])$. We use $\Delta s_f/R = 1.2$, which is typical for metals. *Phase fraction:* Because all molar volumes are identical by assumption, identical numerical values are found for the molar fraction and the volume fraction of the phases. Thus, for solid solvent, $p = N^S/N = V^S/V$, with $p = 1-x$ and $p = 1-x/x^L$ for solid-solid and solid-liquid coexistence respectively. *Interfacial area*: Defining the dimensionless parameter $a$ so that $A = a\pi D^2/4$, we used the empirical function $a = 2^{1/3} q^{2/3} + (1 - 2^{1/3})q$, with $q = 4p(1-p)$, which approximates the numerically exact $A(p)$ of a spherical cap meeting the particle surface at right angles to within better than $\Delta A/A = 7 \times 10^{-3}$ at all $p$. *Capillary energy*: Since $N = \pi D^3/(6\Omega N_A)$, where $\Omega$ and $N_A$ denote the atomic volume and Avogadros number, the area per mole of alloy is $A/N = 3a\Omega N_A/(2D)$. Formally, $\gamma$ can always be related to $\Delta h_f$ through $\gamma = c \Delta h_f / (\Omega^{2/3} N_A)$ with $c$ a dimensionless constant.[17] In units of $\Delta h_f$, the interfacial free energy per mole is therefore $\gamma A/N = 3 a(p) c \Omega^{1/3}/(2D)$. This shows that the size- and materials-dependence of the capillary energy enters the computation by a single dimensionless parameter, the product $c\Omega^{1/3}/D$, which is one of the variables of the computation. In order to specify meaningful values for $D$ we used $\Omega = 0.015$ nm$^3$, typical of metals; typical values of $c$ in metals are $c \approx 0.5$ for solid-liquid interfaces[17] at $T_f$ and $c \approx 1.3$ for grain boundaries.[18] Since we ignore changes in $\gamma$, we used an average value for all interfaces, $c = 0.9$.

### Width of the discontinuous melting interval

Let $T_0$ be the temperature where $G^{S_1}$, $G^{S_2}$, and $G^L$, have a common tangent (which meets the $G^\alpha$ at the solute fractions $x_0^\alpha$) at finite $N$, and define the chemical potentials so that the tangent is horizontal. Near their minima at $T_0$, the $G^\alpha$ are approximated by the series expansions

$$G^\alpha(T,x) = c^\alpha N(T_0 - T) + \frac{1}{2} N(x - x_0^\alpha)^2 / \chi_0^\alpha, \qquad (A1)$$

where the $c^\alpha$ are expansion coefficients and the $\chi_t^\alpha$ denote alloy susceptibilities, $\chi^{-1} = N^{-1} \partial^2 G/\partial x^2$, evaluated at $T_0, x_0^\alpha$. At large $N$ the capillary terms are small, and consequently the $\tilde{G}^{SL}$ in plots such as Fig. 1b) vary much less with $x$ than do the the $G^\alpha$. The equilibrium compositions of solid and liquid in the two-phase state will then be near the eutectic compositions $x^\alpha_e$ of the bulk alloy, and the temperature will be near the bulk eutectic temperature $T_e$. We can therefore approximately take $x^L = x^L_e$, $x^S = x^S_e$, $T_0 = T_e$. With Eq. (5) the free energy for solid-liquid coexistence at small solid fraction $p$ is then

$$G^{SL} = (pc^S + (1-p)c^L)(T_e - T)N + \Gamma(pNv^S)^{2/3}. \qquad (A2)$$

with $p = (x^L_e - x)/(x^L_e - x^S_e)$. The evaluation of $x_0$ requires solving $G^L(T,x) = G^{SL}(T,x)$ for $x$. By means of Eqs. (A1) and (2) this condition can be expressed as

$$\frac{\frac{1}{2} N p^2 (x^L_e - x^S_e)^2 / \chi^L_e}{pN(c^S - c^L)(T - T_e) + p^{2/3} \Gamma (Nv^S)^{2/3}} . \qquad (A3)$$

The two sides of this equation are represented graphically by the two curves, $G^L$ and $G^{SL}$, respectively, in Fig. 1b). It is seen that for large systems, where the point of intersection is near $x = x_e$, the second term on the right-hand side of Eq. (A3) is dominant over the linear term because the slope of $G^{SL}$ diverges as this curve meets $G^L$. In the limit of sufficiently large particles it is therefore allowed to neglect the linear term. There is then a simple solution of Eq. (A3) in terms of the phase fraction $p_0$ at the point of intersection between $G^L$ and $G^{SL}$,

$$p_0 = \left( \frac{2\Gamma \chi_e^L}{(x_e^L - x_e^S)^2} \right)^{3/4} \sqrt{v^S} \, N^{-1/4}. \tag{A4}$$

This provides $x_0$ via Eq. (2). As Eq. (A4) is independent of $T$, it applies in particular at the point of intersection of the three curves $G^{SS}$, $G^{SL}$, and $G^L$, which determines $\Delta x_d$.

**Figures**

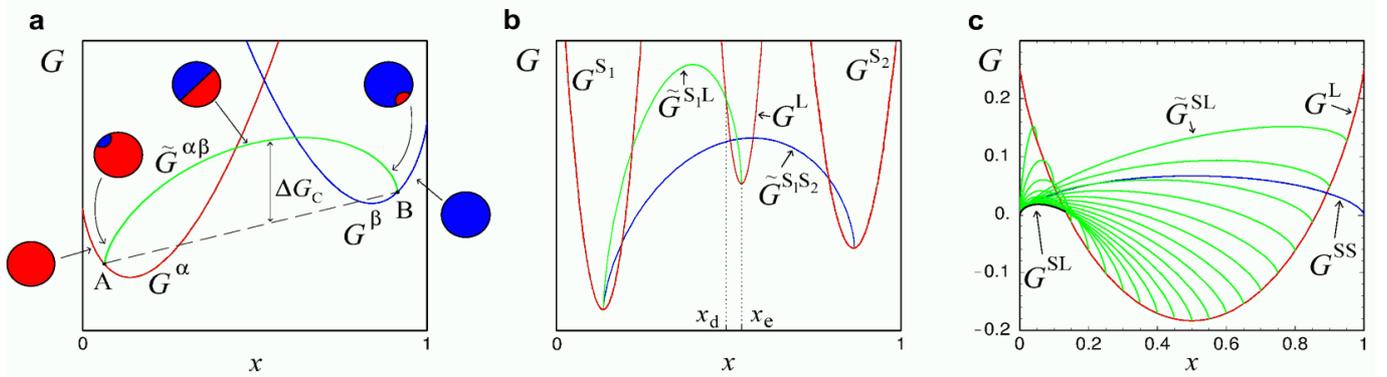

**Figure 1:** Molar free energy functions for finite-size alloys. (**a**) Schematic diagram showing the construction of the free energy curve $\tilde{G}^{\alpha\beta}$ at coexistence of two phases α and β with the arbitrarily selected compositions represented by points A and B. $G^{\alpha}$ and $G^{\beta}$ denote the molar free energies of the single-phase particles. $\Delta G_C$ represents the deviation from the linear behavior. Inserts: schematic cross-sections through a particle. (**b**) Schematic free-energy diagram for two solid phases $S_1$ and $S_2$ and the liquid L slightly above the eutectic temperature of the particle. $x_e$: eutectic composition; $x_d$: limit of discontinuous melting interval. (**c**) Examples from the numerical computation; parameters are $T/T_f = 0.75$ and $D = 5$ nm. $G^{SL}(T,x)$ is the lower envelope of the set of functions $\tilde{G}^{SL}(T,x,x_L)$.

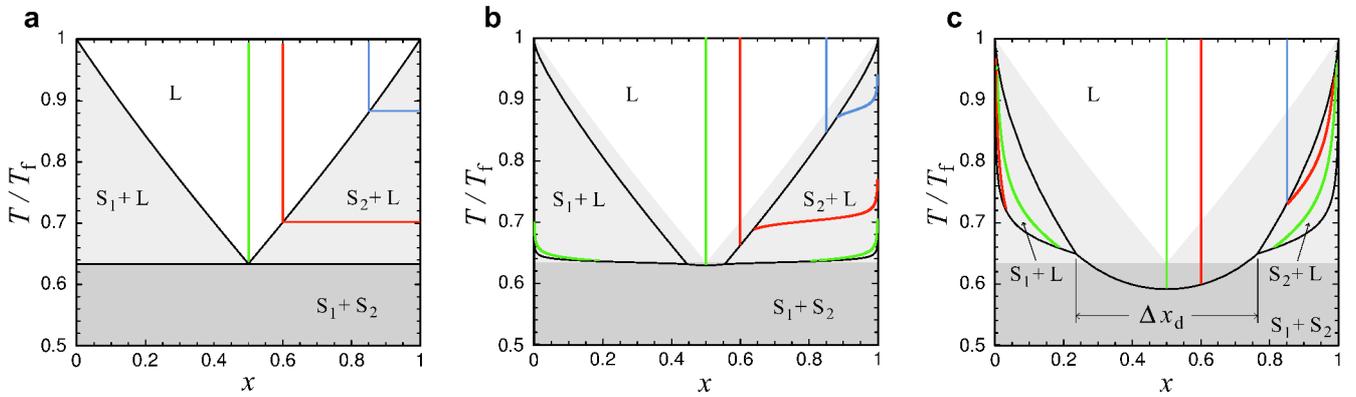

**Figure 2:** Computed alloy phase diagrams. Black: phase coexistence lines; colored: lines of equal solute fraction $x^L$ in the liquid phase for three arbitrarily chosen values of $x^L$. Macroscopic (**a**) and finite particle size $D = 50$ nm (**b**) and $D = 5$ nm (**c**). $\Delta x_d$: discontinuous melting interval. Shaded regions represent phase fields of the macroscopic alloy.